\documentclass[showkeys]{revtex4}

\usepackage{graphicx}
\usepackage{amsmath,amsthm,amssymb,bm,amsfonts}
\usepackage{todonotes}

\usepackage{color}
\usepackage[makeroom]{cancel}
\usepackage[normalem]{ulem}

\def\qbar{{\overline{q}}}

\def\qbold{{\mathbf{q}}}
\def\bbold{{\mathbf{b}}}

\def\fprime{{f^\prime}}

\def\Pcal{{\cal{P}}}

\def\eto{{\rm{e}}}

\newcommand{\tD}{\tilde{D}}

\def\Dbar{\bar{D}}

\newcommand{\bea}{\begin{eqnarray}}
\newcommand{\eea}{\end{eqnarray}}
\newcommand{\be}{\begin{equation}}
\newcommand{\ee}{\end{equation}}

\begin{document}
\vspace*{1.5cm}
\title{\bf Momentum sum rule and factorization of double parton distributions}

\author{Krzysztof Golec-Biernat}
\email{golec@ifj.edu.pl}
\affiliation{Institute of Nuclear Physics Polish Academy of Sciences, 31-342 Cracow, Poland}

\author{Anna M. Sta\'sto}
\email{ams52@psu.edu}
\affiliation{Penn State University, University Park, Pennsylvania 16802, United States}

\begin{abstract}
We show that the momentum sum rule is a necessary condition for factorization of double parton distributions
 into a product of two single parton distributions for small values of the parton momentum fractions $x$ 
 and large enough values of the evolution scale $Q^2$.  This is a somewhat surprising result since the  momentum sum rule involves integration over all values of  the momentum fraction. In essence, the momentum sum rule 
provides a proper relation between the double and single parton distributions,  which is necessary for the small $x$  factorization at large $Q^2$.

\end{abstract}


\maketitle
\section{Introduction}

Multiparton interactions play an important role in high energy scattering of hadrons \cite{Kirschner:1979im,Shelest:1982dg,Zinovev:1982be,Ellis:1982cd,Paver:1982yp, Paver:1983hi, Paver:1984ux, Bukhvostov:1985rn, Kulesza:1999zh,
Snigirev:2003cq,Korotkikh:2004bz,Gaunt:2009re,Blok:2010ge,Ceccopieri:2010kg,Diehl:2011tt,
Gaunt:2011xd,Ryskin:2011kk,Blok:2011bu, Kom:2011nu, Diehl:2011yj,Luszczak:2011zp, 
Manohar:2012jr,Ryskin:2012qx,Gaunt:2012dd,Blok:2013bpa,Broniowski:2013xba, Diehl:2014vaa,vanHameren:2014ava, Golec-Biernat:2014bva, Ceccopieri:2014ufa, Maciula:2014pla, Snigirev:2014eua,  Golec-Biernat:2014nsa, Gaunt:2014rua, Harland-Lang:2014efa, Blok:2014rza, Maciula:2015vza, Golec-Biernat:2015aza, Diehl:2015bca, Broniowski:2016trx, 
Blok:2016lmd, Rinaldi:2016jvu,Golec-Biernat:2016vbt,Ceccopieri:2017oqe, Diehl:2017kgu, Buffing:2017mqm, Blok:2017alw, Diehl:2017wew,  Elias:2017flu,Rinaldi:2018slz, Rinaldi:2018bsf,   Gaunt:2018eix, Diehl:2018wfy, Diehl:2019rdh, Broniowski:2019rmu, 
Diehl:2020xyg, Bali:2020mij, Cabouat:2020ssr, Diehl:2021wpp, Diehl:2021wvd}.  In particular, the double parton scattering (DPS)  in which two pairs of partons from colliding  hadrons take part in a hard scattering process is of special importance. 
The DPS processes  were first observed  at the Tevatron \cite{Akesson:1986iv,Abe:1997bp,Abe:1997xk,Abazov:2009gc} and  are presently studied at the Large Hadron Collider by the  ATLAS \cite{Aad:2013bjm,Aad:2014rua,ATLAS:2014ofp,ATLAS:2016rnd,ATLAS:2016ydt,ATLAS:2018zbr},  CMS \cite{Chatrchyan:2013xxa, CMS:2016liw, CMS:2019jcb, CMS:2022pio} and LHCb \cite{LHCb:2011kri,LHCb:2012aiv,LHCb:2015wvu,LHCb:2016wuo} collaborations.

 The computation of DPS cross sections within the collinear framework makes use of the double parton distribution functions (DPDFs) which  obey QCD evolution equations 
\cite{Kirschner:1979im,Shelest:1982dg,Zinovev:1982be,Snigirev:2003cq,Korotkikh:2004bz,Ceccopieri:2010kg,Ceccopieri:2014ufa, Gaunt:2009re}, 
similar to the Dokshitzer-Gribov-Lipatov-Altarelli-Parisi (DGLAP)  evolution equations for the single parton distribution functions (PDFs), 
see also \cite{Ryskin:2012qx} for a pedagogical presentation and \cite{Diehl:2017wew} for an overview.
The evolution equations for DPDFs conserve sum rules which relate the double and single parton distribution functions.  
It means that once  these rules are imposed on initial conditions for the evolution equations at an initial  scale $Q^2_0$,  they are also obeyed by evolved distributions at the scale $Q^2 \ge Q^2_0$.

All the attempts  to  construct  conditions which exactly satisfy the new sum rules  were rather unsuccessful 
until now; see, e.g., 
Refs.~\cite{Korotkikh:2004bz,Gaunt:2011xd,Golec-Biernat:2014bva} with an exception of the analysis  \cite{Broniowski:2013xba} for valence quarks only.  Also in a pure gluonic case, the  double gluon distribution $D_{gg}$  was proposed in Ref.  \cite{Golec-Biernat:2015aza} which obeys the momentum sum rule relating $D_{gg}$ and the single gluon distribution $D_g$. This was achieved due to a particular form of $D_g$, used in   global fits to hard scattering data as an initial condition for the DGLAP evolution equations. The parameters of $D_g$ fully determined the parameters of the initial $D_{gg}$ such that the momentum sum rule is fulfilled.

In most practical applications of DPDFs it is usually assumed  that  for small parton momentum fractions, $x_{1},x_2\ll 1$, the DPDFs factorize into a product of single PDFs, e.g. $D_{gg}(x_1,x_2,Q^2)\approx D_g(x_1,Q^2) D_g(x_2,Q^2)$. 
However, the example of the initial condition  for $D_{gg}$ from \cite{Golec-Biernat:2015aza} shows that the small $x$ factorization can be strongly violated at the initial scale $Q_0^2=1~{\rm GeV}^2$ but   is restored after the evolution to a sufficiently large $Q^2$. 

Therefore, it is a main goal of this paper to elucidate the issue of the small $x$ factorization
in the framework of the QCD evolution equations  for DPDFs. In particular, we will show that the momentum sum rule,
which in principle must be imposed on the initial distributions for the QCD evolution equations, is a necessary condition for the small $x$ factorization, i.e., without the momentum sum rule the small $x$  factorization is violated.
This result has important phenomenological consequences for the current prescriptions for the  initial conditions for DPDFs which use the information on the single PDFs. Such parton distributions usually fulfill the momentum sum rule only approximately and must be used with some care in precise studies based on the QCD evolution equations.

It should be mentioned that the issue of the small $x$ factorization is studied here for the DPDFs with transverse momentum
$\qbold=0$, when the momentum sum rule is valid. However, there are strong indications that for $\qbold\ne 0$ there is no factorization at all due to correlations  between partons in the impact parameter space with the variable $\bbold$, being Fourier conjugate to $\qbold$ \cite{Diehl:2014vaa,Rinaldi:2016jvu, Diehl:2020xyg}. Nevertheless, our studies are important since the form of the DPDFs at 
$\qbold=0$ is essential for the generalization to the $\qbold\ne 0$ case.

The paper is organized as follows. In Sec.~II we present main information about the DPDFs,  their QCD evolution equations  and  sum rules. In Sec.~III we restrict  ourselves to the pure gluon case with the double gluon distribution $D_{gg}$ for a detailed numerical analysis presented in the next sections. In Sec.~IV, based on a simple ansatz for the single and double gluon distributions,  we present two numerical examples showing  that the small $x$ factorization of $D_{gg}$ is conserved or restored in the QCD evolution. In Sec.~V we illustrate on  numerical examples  that the momentum sum rule is a necessary condition for the latter conclusions. In Sec.~VI we provide an analytical insight to these results, using the Mellin moment representation. Finally, in Sec.~VII we present the current situation concerning the relation between the double and single parton distributions, including also quarks into the considerations.
In Appendix A the proof of  the momentum sum rule conservation by the evolution equations is presented, while in Appendix B the asymptotic solution to the evolution equations is discussed in the Mellin moment space.

\section{Evolution equations and sum rules}
\label{sec:2}

The double parton distributions $\Dbar_{f_1f_2}(x_1,x_2,\bbold)$ have probabilistic interpretation of the number density of pairs of partons with longitudinal momentum fractions $x_1$ and $x_2$ (for which $0< x_1+x_2 \le 1$) at a relative transverse vector $\bbold$ \cite{Diehl:2011tt,Diehl:2011yj}. Parton flavours (including gluon) are denoted by $f_1$ and $f_2$.
In this paper, we are  interested in the double parton distributions integrated over $\bbold$, which are equal to the 
Fourier transform,
\be
D_{f_1f_2}(x_1,x_2,\qbold) = \int d^2\bbold\,\eto^{i\qbold\cdot\bbold}\,\Dbar_{f_1f_2}(x_1,x_2,\bbold)\;,
\ee
taken at $\qbold=0$,
\be\label{eq:2new}
D_{f_1f_2}(x_1,x_2,\qbold=0) = \int d^2\bbold\, \Dbar_{f_1f_2}(x_1,x_2,\bbold)\;.
\ee
For the physical meaning of the transverse momentum $\qbold$, see \cite{Diehl:2011yj}. 
The  DPDFs (\ref{eq:2new}) obey QCD evolution equations  with respect to two hard scales, $Q_1^2$ and $Q^2_2$, which are  also present due to renormalization. We consider the DPDFs with equal hard scales, $Q_1^2=Q_2^2\equiv Q^2$, i.e.
\be
\label{eq:dpdfdef}
D_{f_1f_2}(x_1,x_2,Q^2)\,\equiv\,D_{f_1f_2}(x_1,x_2,\qbold=0,Q^2,Q^2)\; .
\ee
Introducing the evolution parameter 
\be
\label{eq:tparam}
t=t(Q^2)=\int^{Q^2}_{Q_0^2} \frac{\alpha_s(\mu^2)}{2\pi}\frac{d\mu^2}{\mu^2}
=\frac{6}{33-2n_f}\ln\frac{\ln(Q^2/\Lambda_{\rm QCD}^2)}{\ln(Q_0^2/\Lambda_{\rm QCD}^2)} \; ,
\ee
where $\alpha_s(\mu^2)$ is the  running  strong coupling constant in the leading order (LO) approximation,  the evolution equations  read \cite{Shelest:1982dg,Zinovev:1982be,Snigirev:2003cq,Gaunt:2009re,Ceccopieri:2010kg,Ryskin:2011kk}
\begin{align}\nonumber
\label{eq:twopdfeq}
\frac{\partial}{\partial t}\, D_{f_1f_2}(x_1,x_2,t)
=\sum_{f'}
\Bigg\{&\int^{1-x_2}_{x_1} \frac{du}{u} \,{{P}}_{f_1f'}\!\left(\frac{x_1}{u}\right) D_{f' f_2}(u,x_2,t)
+\int_{x_2}^{1-x_1} \frac{du}{u}\,{{P}}_{f_2f'}\!\left(\frac{x_2}{u}\right)D_{f_1f'}(x_1,u,t)
\\
&+ \frac{1}{x_1+x_2}\,{P}_{f'\to f_1f_2}\!\left(\frac{x_1}{x_1+x_2}\right) D_{f'}(x_1+x_2,t)\Bigg\} \; .
\end{align}
Here, the functions  ${P}$ on the rhs are  the  LO Altarelli-Parisi splitting functions (with virtual corrections included) and the summation is performed over quark/antiquark flavors and gluon.  
The third, {\it splitting term}  on the rhs corresponds to the splitting of one parton into two daughter partons,  described by the LO Altarelli-Parisi splitting function for the real emission, $P_{f'\to f_1f_2}=P^{R}_{f^\prime f_1}$. It also contains the single PDFs; thus Eq.~\eqref{eq:twopdfeq} has to be solved together with the ordinary DGLAP equations
\be
\label{eq:dglap}
\frac{\partial}{\partial t}\, D_{f}(x,t) =\sum_{f'}
\int^{1}_{x} \frac{du}{u} \,{{P}}_{ff'}\!\left(\frac{x}{u}\right) D_{f' }(u,t) \; .
\ee
To formulate the next-to-leading order (NLO) evolution equations for the DPDFs, the two-loop  splitting functions
${P}_{f'\to f_1f_2}$ were calculated in \cite{Diehl:2019rdh,Diehl:2021wpp}. The NLO formulation, however, is beyond the scope of this paper.

The DPDFs for $\qbold=0$ obey sum rules which can be derived starting from the definition of  DPDFs based on
the light-cone nucleon wave function \cite{Gaunt:thesis}. What is most important, these sum rules are consistent with the evolution equations, which means
that once the sum rules are assumed at the initial scale $t=0$, they are conserved during the QCD evolution to any $t$ 
\cite{Gaunt:2009re,Diehl:2011tt,Gaunt:2011xd}.
These are: the momentum sum rule 
\begin{eqnarray}
\label{eq:momrule1}
\sum_{f_1}\int_{0}^{1-x_2}dx_1\,x_1D_{f_1f_2}(x_1,x_2,t)=(1-x_2)D_{f_2}(x_2,t) \; ,
\end{eqnarray}
and the valence  quark number sum rule 
\begin{eqnarray}
\label{eq:valrule1}
\int_0^{1-x_2}dx_1\!\left\{D_{q_if_2}(x_1,x_2,t)-D_{\qbar_i f_2}(x_1,x_2,t)\right\} =
(N_{q_i}-\delta_{q_if_2}+\delta_{\qbar_i f_2}) D_{f_2}(x_2,t)\,,
\end{eqnarray}
where $q_i=u,d,s$ and  $N_u=2, N_d=1,N_s=0$ are the corresponding valence quark numbers.  
Their form can be readily understood from the probability theory, treating the ratios of the double to single distributions
as conditional probabilities, see \cite{Gaunt:2009re} for details. Analogous sum rules 
also hold with respect to the second parton momentum $x_2$. To ensure this property, the parton exchange symmetry must be imposed on the initial conditions, which is kept by the evolution to any $t$,
\be
 D_{f_1f_2}(x_1,x_2,0) = D_{f_2f_1}(x_2,x_1,0) \; .
 \ee
 Relations (\ref{eq:momrule1}) and (\ref{eq:valrule1}) should be considered together with the momentum sum rule for single PDFs,
\begin{align}
\label{eq:singlemom}
\sum_f\int_0^1dx\,xD_f(x,t)=1 \; ,
\end{align}
and the valence quark number sum rule 
\begin{align}
\label{eq:singleval}
\int_0^1dx\,\left\{D_{q_i}(x,t)-D_{\qbar_i}(x,t)\right\}=N_{q_i}\,,
\end{align}
which are also conserved by the DGLAP evolution equations. 
 In Appendix A we present the proof of the conservation of the momentum sum rule (\ref{eq:momrule1})  by the evolution equations 
(\ref{eq:twopdfeq}) and (\ref{eq:dglap}),
in which the presence of the splitting term in Eq.~(\ref{eq:twopdfeq}) plays the crucial role.

\section{Pure gluonic case}

In the following, we shall consider only the gluonic case for the purposes of simplicity of the presentation. 
In this case, we only deal with the double gluon distribution $D_{gg}$ and the single gluon distribution $D_g$,
which obey the evolution equations (\ref{eq:twopdfeq}) reduced to the gluon sector, 
\begin{align}\nonumber
\label{eq:tevggl}
\frac{\partial}{\partial t}\, D_{gg}(x_1,x_2,t)
=
&\int^{1-x_2}_{x_1} \frac{du}{u} \,{{P}}_{gg}\!\left(\frac{x_1}{u}\right) D_{gg}(u,x_2,t)
+\int_{x_2}^{1-x_1} \frac{du}{u}\,{{P}}_{gg}\!\left(\frac{x_2}{u}\right)D_{gg}(x_1,u,t)
\\
&+ \frac{1}{x_1+x_2}\,P^R_{gg}\!\left(\frac{x_1}{x_1+x_2}\right) D_{g}(x_1+x_2,t) \; ,
\end{align}
and
\be
\label{eq:dglapgl}
\frac{\partial}{\partial t}\, D_{g}(x,t) =\int_x^1 \frac{dz}{z}\,{P}_{gg}\!\left(\frac{x}{z}\right) D_g(z,t) \; .
\ee 
The momentum sum rules (\ref{eq:momrule1}) and (\ref{eq:singlemom}), conserved by the above evolution equations,
take the following form
\be
\label{eq:momdglue}
\int_{0}^{1-x_2}dx_1\,x_1D_{gg}(x_1,x_2,t)=(1-x_2)D_{g}(x_2,t) \; ,
\ee
and
\be
\label{eq:momsglue}
\int_0^1dx\,xD_g(x,t)=1 \; .
\ee
With the normalization to one in the last condition, we assume that gluons carry all the proton longitudinal momentum. 
The momentum momentum sum rules should be valid for any $t$, including $t=0$ where the initial conditions for the evolution equations (\ref{eq:tevggl}) and (\ref{eq:dglapgl}) are specified. As in the general case, we assume the gluon exchange symmetry for the initial conditions to ensure the momentum sum rule (\ref{eq:momdglue}) with respect to the second momentum fraction,
\be
D_{gg}(x_1,x_2,0)=D_{gg}(x_2,x_1,0) \; .
\ee

The conservation of the momentum sum rules by the evolution equations means that the single  gluon distribution obtained from  (\ref{eq:momdglue}), 
\be\label{eq:12new}
D_g(x_2,t)=\frac{1}{1-x_2} \int_{0}^{1-x_2}dx_1\,x_1D_{gg}(x_1,x_2,t) \; ,
\ee
obeys  equation  (\ref{eq:dglapgl}) for any $t$, including $t=0$. This imposes strong constraint on the double gluon distribution alone
which results from the momentum sum rule (\ref{eq:momsglue}), 
\be
\label{eq:13new}
\int_0^1dx_2\int_{0}^{1-x_2}dx_1\,\frac{x_1x_2 }{1-x_2}\,D_{gg}(x_1,x_2,t)=1\;.
\ee

In the next section, we will consider a simple ansatz for the initial gluon distributions which satisfies the momentum sum rules (\ref{eq:momdglue}) and (\ref{eq:momsglue}), which will  allow for detailed studies of the relation between the momentum sum rule and  the small $x$ factorization of $D_{gg}(x_1,x_2,t)$.

\section{A simple example}

\label{sec:simple}

\begin{figure}[t]
\begin{center}
\includegraphics[width=0.55\textwidth]{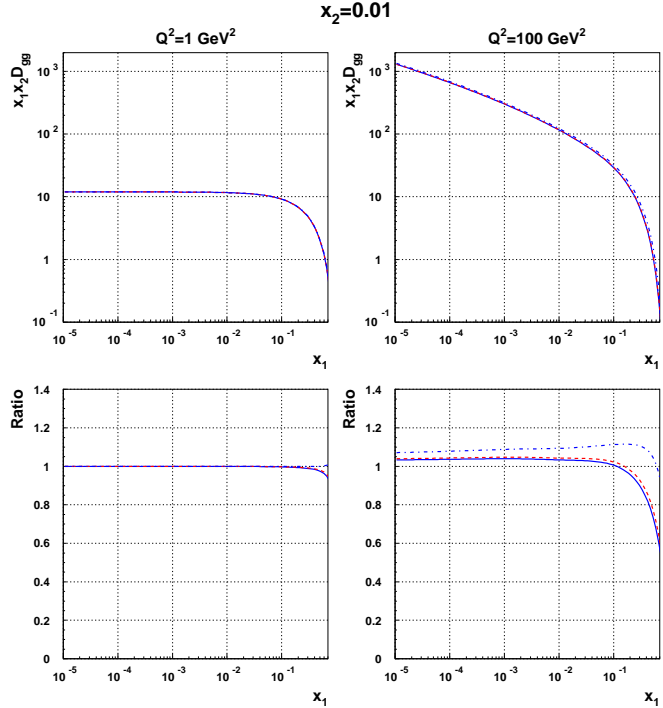}
\caption{In the upper plots: $x_1x_2D_{gg}(x_1,x_2)$ as a function of $x_1$ at the  initial scale $Q^2=1~{\rm GeV}^2$ (left panels) and final scale 
$Q^2=100~{\rm GeV}^2$ (right panels) and fixed $x_2=10^{-2}$ for  model (\ref{eq:4}) (solid lines), Gaunt model (\ref{eq:6}) (dashed lines) 
and fully factorized ansatz (\ref{eq:14new}) (dash-dotted lines). The parameters in (\ref{eq:4}) are: $\alpha_g=-1$ and $\beta_g=2.5$. 
In the bottom plots: the corresponding ratios (\ref{eq:17}). 
Both the momentum sum rule (\ref{eq:momdglue}) and the small $x$ factorization (\ref{eq:18a}) 
hold true for our model.
}
\label{fig:1}
\end{center}
\end{figure}

A general construction for the  initial conditions for the single and double gluon distributions which obey the sum rules was proposed in \cite{Golec-Biernat:2015aza}.
This framework is based on sums over Dirichlet distributions, where relations between the powers and the normalizations can be found to ensure that the sum rules for double and single distributions are simultaneously satisfied. In the simple example below we shall follow the construction of Ref.~\cite{Golec-Biernat:2015aza} for the simple case of the distributions having one term in the sum.

Let us consider the single gluon distribution  of the form
\be\label{eq:1}
D_g(x,0) = A_g \,x^{\alpha_g} (1-x)^{\beta_g} \; ,
\ee
where the normalization constant is determined from the momentum sum rule (\ref{eq:momsglue}) 

\be
\label{eq:1a}
A_g \int_0^1 dx \,x^{\alpha_g+1} (1-x)^{\beta_g} = 1\; .
\ee
Using the well-known formula relating the Euler beta function to the  gamma functions,
\be
\int_0^1 dx \, x^\alpha (1-x)^\beta=\frac{\Gamma(\alpha+1)\Gamma(\beta+1)}{\Gamma(\alpha+\beta+2)}\;,
\ee
we obtain
\be
 \label{eq:2}
 A_g=\frac{\Gamma(\alpha_g+\beta_g+3)}{\Gamma(\alpha_g+2)\Gamma(\beta_g+1)} \; .
\ee
It can be shown \cite{Golec-Biernat:2015aza} that the double gluon distribution 
which obeys the momentum sum rule (\ref{eq:momdglue}) with $D_g$ of the form (\ref{eq:1}) is given by
\be\label{eq:4}
D_{gg}(x_1,x_2,0) = A_{gg}  \, (x_1 x_2)^{\alpha_g} (1-x_1-x_2)^{\beta_g-\alpha_g-1} \; ,
\ee
where the normalization constant
\be\label{eq:24new}
A_{gg} = \frac{\Gamma(\beta_g+2)}{\Gamma(\alpha_g+2)\Gamma(\beta_g-\alpha_g)} \, A_g \; .
\ee
This is a remarkable result since 
the parameters of the double distribution $D_{gg}$ are completely determined by the parameters of 
the single distribution $D_g$ through the momentum sum rule (\ref{eq:momdglue}). Note also, that the powers governing the small $x_{1,2}$ behavior are the same in Eqs.~\eqref{eq:1} and \eqref{eq:4}.

 \begin{figure}[t]
\begin{center}
\includegraphics[width=0.55\textwidth]{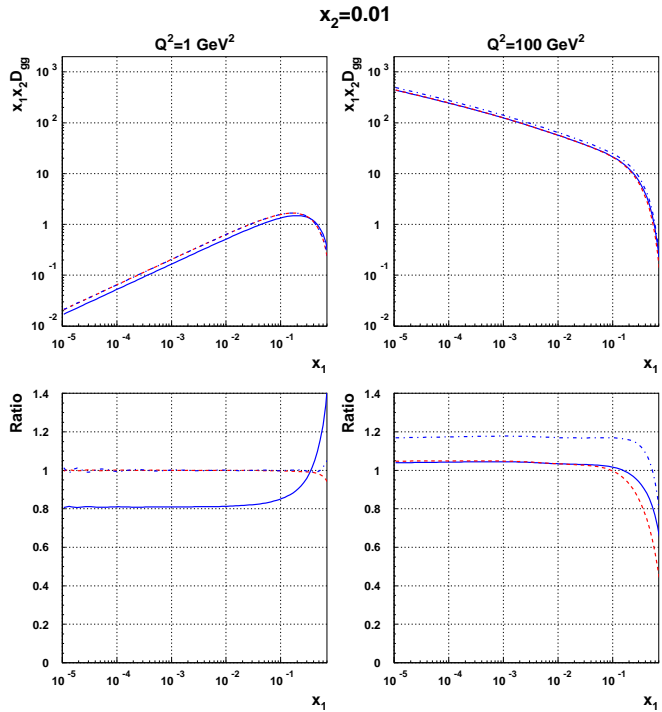}
\caption{The same as in Fig.~\ref{fig:1} but for $\alpha_g=-0.5$ and $\beta_g=2.5$  in (\ref{eq:4}). The momentum sum rule is  valid but the  small $x$ factorization is violated by the initial conditions (solid line in the left bottom plot).  
After the  evolution to large $Q^2=100 \, \rm GeV^2$, the small $x$ factorization is to a good approximation restored (solid line on the right bottom plot).
}
\label{fig:3}
\end{center}
\end{figure}

For  $\alpha_g=-1$ we have $A_{gg} = A_g^2$  and the double gluon distribution factorizes for  $x_1,x_2\ll 1$, 
\be\label{eq:18a}
D_{gg}(x_1,x_2,0) \approx  A_g^2\, (x_1 x_2)^{\alpha_g}\,\approx D_g(x_1,0) D_g(x_2,0) \;.
\ee
We illustrate this case in Fig.~\ref{fig:1} choosing $\alpha_g=-1$ and $\beta_g=2.5$ in Eq.~(\ref{eq:1})  for the  single gluon distribution  
defined at the initial scale $Q_0^2=1~{\rm GeV}^2$.  The choice of $\alpha_g=-1$ corresponds to ``flat"  initial distribution $xD_g(x,0)$ at small values of $x$.  The corresponding initial double gluon distribution  was computed  from Eq.~(\ref{eq:4}). In the two upper plots we show the distribution $x_1x_2D_{gg}(x_1,x_2,t)$ at the initial scale and final scale equal to $Q^2=100~{\rm GeV}^2$ (solid lines). Both plots are presented in the way
appropriate for the studies of the small $x$ limit of the double gluon distributions. Namely, we plot them as functions of $x_1$ for fixed $x_2=10^{-2}$ to study the limit $x_1\to 0$. 
The two bottom plots show the ratio
\be\label{eq:17}
{\rm Ratio}=\frac{D_{gg}(x_1,x_2,t)}{D_g(x_1,t)D_g(x_2,t)} \; ,
\ee
at the corresponding values of $Q^2$ [related to $t$ by Eq.~(\ref{eq:tparam})]. We see that both before and after the evolution, the small $x$ factorization holds for the double gluon distribution to a good approximation  up to $x_1=10^{-1}$.

 We also show the results obtained for the Gaunt-Stirling (GS) prescription \cite{Gaunt:2009re} for the initial condition (dashed lines),
 \be\label{eq:6}
D_{gg}(x_1,x_2,0)=D_g(x_1,0) D_g(x_2,0)\frac{(1-x_1-x_2)^2}{(1-x_1)^2(1-x_2)^2} \; ,
\ee
 as well as for the fully factorized ansatz  (dash-dotted lines)
 \be
\label{eq:14new}
D_{gg}(x_1,x_2,0) = D_g(x_1,0)\,D_g(x_2,0) \,\theta(1-x_1-x_2)\;,
\ee
where the single gluon distribution (\ref{eq:1})  with the assumed parameters was used.
Both prescriptions give the approximate factorization
 after the evolution to $Q^2=100~{\rm GeV}^2$, although the result for the fully factorized ansatz is  slightly worse. This agreement can be attributed to 
a rather weak violation of the momentum sum rule by these distributions. 
In particular, the integral (\ref{eq:13new}) computed for the ansatz (\ref{eq:6}) gives $1.04$ 
while for the  ansatz (\ref{eq:14new}) we find $1.3$ (instead of 1 as in our model).

For $\alpha_g\ne -1$,  the small $x$ factorization of the initial double gluon distribution (\ref{eq:4})  is no longer true,
\be
D_{gg}(x_1,x_2,0) \ne  D_g(x_1,0) D_g(x_2,0)\;,
\ee
 but the momentum sum rule is still fulfilled.
This means that the momentum sum rule is neither necessary nor sufficient condition for the small $x$ factorization of the initial double gluon distribution.  We illustrate this fact in Fig.~\ref{fig:3} with  the initial  double gluon distribution (\ref{eq:4})  with the parameters 
$\alpha_g=-0.5$ and $\beta_g=2.5$ (solid line  on the left bottom plot).   However, it is very interesting that  the  small $x$ factorization of $D_{gg}(x_1,x_2)$ is to a good approximation restored by the evolution to $Q^2=100~{\rm GeV}^2$ 
since the ratio (\ref{eq:17})  is close to one (solid line on the right bottom plot).

 The same effect is also  clearly visible in the analysis in Ref.~\cite{Golec-Biernat:2015aza} with a realistic ansatz
 based on a  particular form of the single PDFs in the MSTW08 parametrization \cite{Martin:2009iq},
\be
\label{eq:26abc}
D_f(x,0)=\sum_{i=1}^3 A_f^i \,x^{\alpha_f^i}(1-x)^{\beta_f^i} \; ,
\ee
where $f$ denotes quark/antiquark flavor or gluon.
In the pure gluonic case, the double gluon distribution is fully determined by the parameters
of the single gluon distribution $D_g(x,0)$, using the momentum sum rule  (\ref{eq:momdglue}),
\be\label{eq:47}
D_{gg}(x_1,x_2,0)= \sum_{i=1}^3 A_{gg}^i \,(x_1x_2)^{\alpha_g^i}(1-x_1-x_2)^{\beta_g^i-\alpha_g^i-1} \; ,
\ee
where $A_{gg}^i$ is also given in terms of the parameters $\alpha_g^i$ and $\beta_g^i$.  
The initial condition (\ref{eq:47}) strongly violates the small $x$ factorization at the scale $Q_0^2=1~{\rm GeV}^2$.
However, after the evolution to $Q^2=100~{\rm GeV}^2$ the small $x$ factorization is restored, see 
in Fig.~1 in   Ref.~\cite{Golec-Biernat:2015aza}. 
In the next section, we will explore this phenomenon in more detail.

\section{Momentum sum rule violation}

 \begin{figure}[t]
\begin{center}
\includegraphics[width=0.55\textwidth]{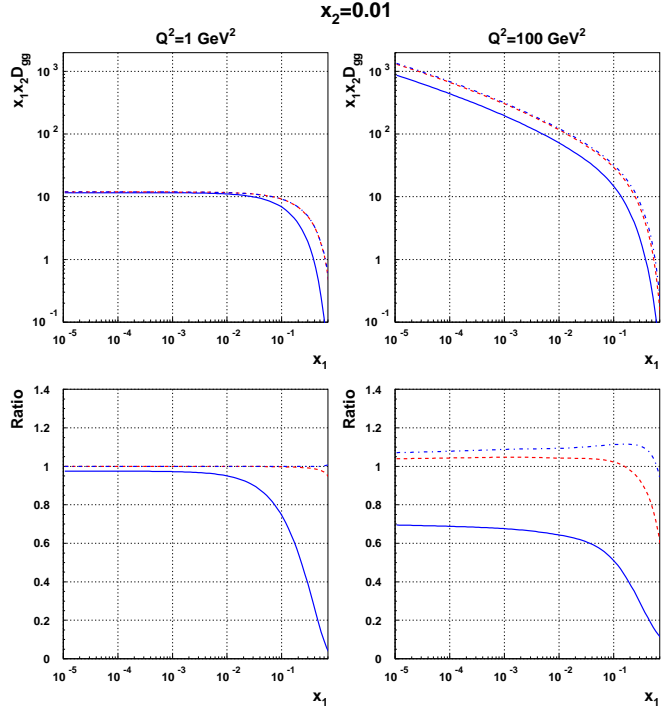}
\caption{The same as in Fig.~\ref{fig:1} but for $\eta=5$ in the initial condition (\ref{eq:31new}) (solid lines). The momentum sum rule is no longer valid but  the  small $x$ factorization    holds true
(solid line on the left bottom plot), while it is violated after the evolution (solid line on the right bottom plot).
}
\label{fig:2}
\end{center}
\end{figure}

In order to understand the role of the momentum sum rule conservation for the  restoration of the small $x$ factorization, 
we consider  the single gluon distribution (\ref{eq:1}) with normalization (\ref{eq:2}),
\be\label{eq:32n}
D_g(x,0) = 
A_g\, x^{\alpha_g}(1-x)^{\beta_g}\;,
\ee
while the double gluon distribution
(\ref{eq:4}) with normalization (\ref{eq:24new}) is modified by changing the parameter $\eta=\beta_g-\alpha_g-1$, which controls the large $x$ behaviour, i.e., 
\be\label{eq:31new}
D^{(m)}_{gg}(x_1,x_2,0) = A_{gg}  \, (x_1 x_2)^{\alpha_g} (1-x_1-x_2)^{\eta}\;,
\ee
where $\eta$ is now  arbitrary. The single gluon distribution still obeys the momentum sum rule (\ref{eq:momsglue}) but  relation (\ref{eq:momdglue}) is violated. Indeed, by computing the  gluon distribution from relation 
(\ref{eq:12new}), 
\be
D_g^{(m)}(x_2,0)=\frac{1}{1-x_2} \int_{0}^{1-x_2}dx_1\,x_1D_{gg}^{(m)}(x_1,x_2,0)\;,
\ee
we obtain
\be
D^{(m)}_g(x,0) =  A_g\,\frac{\Gamma(\beta_g+2)\Gamma(\eta+1)}{\Gamma(\alpha_g+\eta+3)\Gamma(\beta_g-\alpha_g)}\,
x^{\alpha_g}\,(1-x)^{\alpha_g+\eta+1}\;,
\ee
Thus, only for $\eta=\beta_g-\alpha_g-1$ we find $D^{(m)}_g(x,0) = D_g(x,0)$ and the  momentum sum rule (\ref{eq:momdglue})
is fulfilled.  For any other $\eta$, this rule is violated by  the initial distributions $D_{gg}^{(m)}(x_1,x_2,0)$ and $D_g(x,0)$. 

We illustrate this situation in Fig.~\ref{fig:2} for the choice of the parameters $\alpha_g=-1$, $\beta_g=2.5$ and $\eta=5$.
Since $\beta_g-\alpha_g-1=2.5\ne 5$ we deal with the momentum sum rule violation. Nevertheless, the small $x$ factorization
holds at the initial scale, which  is shown on the left bottom plot in Fig. \ref{fig:2} by the solid line, being close to one for $x_1<10^{-2}$.
However, the factorization is strongly violated after  the  evolution to $Q^2=100~{\rm GeV}^2$ (solid line  on the right bottom plot) in the same range of $x_1$, 
 \be\label{eq:33}
\frac{D_{gg}^{(m)}(x_1,x_2,t)}{D_g(x_1,t)D_g(x_2,t)} \approx 0.6-0.7 \; .
\ee
This can be attributed to the mismatch between the single gluon distributions, $D_g(x,t)$ and $D_g^{(m)}(x,t)$. We numerically checked that if $D_g^{(m)}(x,t)$ is substituted to the ratio (\ref{eq:33}), after the additional rescaling of $D^{(m)}_{gg}(x_1,x_2,0)$ such that
the momentum sum rule (\ref{eq:momsglue}) is fulfilled by $D_g^{(m)}(x,0)$, we find  factorization at the 
scale $Q^2=100~{\rm GeV}^2$. The reason for the additional rescaling will become clear from the analytical insight presented in the next section.

To summarize the presented example, the violation of the momentum sum rule (\ref{eq:momdglue}) by the initial conditions
leads to a mismatch between the single  and double gluon distributions, which is manifested in  the  violation of the small
$x$ factorization even at high scales. 
On the other hand, when the momentum sum rules (\ref{eq:momdglue}) and (\ref{eq:momsglue}) are fulfilled,   then the approximate restoration  of the small $x$ factorization is observed  after the  evolution to high scales.

\section{Mellin moment  formulation}

In order to give an analytical insight into the relation between the momentum sum rule and small $x$ factorization, let us introduce the Mellin moments for a single  gluon distribution,
\be
\tD_g(n,t)=\int_0^1 dx \,x^{n-1} D_g(x,t) \; ,
\ee
and for a double gluon distribution,
\be
\tD_{gg}(n_1,n_2,t)=\int_0^1 dx_1 \int_0^1 dx_2 \, x_1^{n_1-1}x_2^{n_2-1} \, \theta(1-x_1-x_2)\,D_{gg}(x_1,x_2,t) \; .
\ee
The momentum sum rule (\ref{eq:momdglue}) in terms of the Mellin moments is given by
\be
\label{eq:mom1mel}
\tD_{gg}(n_1,2,t)=\tD_g(n_1,t)-\tD_g(n_1+1,t) \; ,
\ee
while the momentum sum (\ref{eq:momsglue})  reads as
\be
\label{eq:mom2mel}
\tD_g(2,t)=1 \; .
\ee

It was shown in \cite{Golec-Biernat:2014nsa} that the solution to the evolution equations (\ref{eq:tevggl}) and (\ref{eq:dglapgl}) in the Mellin moment space reads as
\begin{align}
\tD_{gg}(n_1,n_2,t) &= \eto^{\gamma(n_1)t+\gamma(n_2)t} \,\tD_{gg}(n_1,n_2,0) 
+\int_0^t dt' \eto^{\gamma(n_1)(t-t')+\gamma(n_2)(t-t')}\, \tilde{\gamma}(n_1,n_2) \,\tD_g(n_1+n_2-1,t') \; ,
\label{eq:dpdfsolm}
\end{align}
where the gluonic anomalous dimension equals
\be
\gamma(n)=\int_0^1 dx \,x^{n-1} \,{{P}}_{gg}(x) \; ,
\ee
while
\be
\tilde{\gamma}(n_1,n_2) = \int_0^1 dx\,x^{n_1-1}(1-x)^{n_2-1}\,P^R_{gg}(x) \; .
\label{eq:tdgamma}
\ee

In Appendix B we show that in  the small $x$ limit, when $x_1\to 0$ and $x_2={\rm fixed}$ and small (e.g. $x_2 = 10^{-2}$ as in the presented analysis), the solution for $t\to\infty$ is  given in terms of the Mellin moments by
\be\label{eq:37abc}
\tD_{gg}(n_1,n_2,t) \simeq e^{\gamma(n_1)t +\gamma(n_2)t} \left( \tD_{gg}(n_1,n_2,0)+\tD_g(n_1+n_2-1,0)\right) \; .
\ee
where $(n_1-1)\to 0$ and $(n_2-1)$ is finite.

This needs to be compared with the numerically found result which gives us approximate factorization at low $x$
\be\label{eq:38abc}
\tD_{gg}(n_1,n_2,t) \simeq e^{\gamma(n_1)t +\gamma(n_2)t}  \; \tD_{g}(n_1,0)\,\tD_{g}(n_2,0) \; .
\ee
So the small $x$ factorization  holds if we have
\be
\tD_{gg}(n_1,n_2,0)+\tD_g(n_1+n_2-1,0) = \tD_{g}(n_1,0)\,\tD_{g}(n_2,0) \; .
\label{eq:mystery_relation}
\ee
For $n_2=2$ we find the following relation
\be
\tD_{gg}(n_1,2,0)+\tD_g(n_1+1,0) = \tD_{g}(n_1,0)\,\tD_{g}(2,0) \; ,
\ee
Assuming the momentum sum rule for the single gluon distribution, $D_{g}(2,0)=1$, we obtain
\be
\tD_{gg}(n_1,2,0) = \tD_{g}(n_1,0)-\tD_g(n_1+1,0)\;,
\ee
which is  the momentum sum rule (\ref{eq:mom1mel}) to be satisfied by the initial condition.
This is a    necessary condition for the small $x$ factorization  at sufficiently large $t$.  Note  that to arrive at this conclusion, the momentum sum rule (\ref{eq:mom2mel}) has to be fulfilled, which explains the numerical observation presented in the previous section.

\section{General case}

\begin{figure}[t]
\begin{center}
\includegraphics[width=0.55\textwidth]{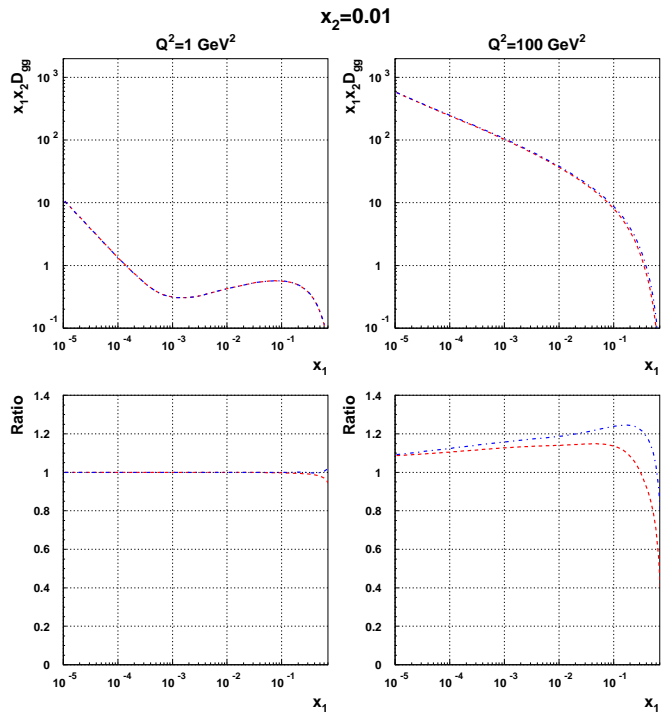}
\caption{In the upper plots: 
the double gluon distributions $x_1x_2 D_{gg}(x_1,x_2)$ as a function of $x_1$ at  $Q^2=1,100~{\rm GeV}^2$ and fixed $x_2=10^{-2}$ for the fully factorized prescription (\ref{eq:44abc}) (dash-dotted lines) and the GS ansatz (\ref{eq:6abc}) (dashed lines) with the initial conditions given by the MSTW08 PDFs. The corresponding ratios (\ref{eq:17}) are shown in the bottom  plots. 
The evolution equations with quarks were used for the results at $Q^2=100~{\rm GeV}^2$.
}
\label{fig:5}
\end{center}
\end{figure}

Let us consider the case with quarks governed by the evolution equations and sum rules presented in  Sec.~\ref{sec:2}.
The momentum sum rules (\ref{eq:momrule1}) and (\ref{eq:singlemom}) lead to the following consistency condition
which should be satisfied by the initial double parton distributions $D_{f_1,f_2}(x_1,x_2,0)$:
\be
\label{eq:32abc}
\sum_{f_1,f_2}\int_0^1dx_2\int_{0}^{1-x_2}dx_1\,\frac{x_1x_2 }{1-x_2}\,D_{f_1f_2}(x_1,x_2,0)=1 \; ,
\ee
where the summation over quark/antiquark flavors and gluon is performed. 
The  departure from 1 of the integral on the lhs is the indication that the momentum sum rule (\ref{eq:momrule1})  is violated.

In Fig.~\ref{fig:5} we analyze the small $x$ factorization for the double gluon distribution $x_1x_2 D_{gg}(x_1,x_2,t)$ obtained from 
the full evolution equations (\ref{eq:twopdfeq})  with the initial conditions given the Gaunt-Stirling (GS)
ansatz (dashed lines)
\be
\label{eq:6abc}
D_{f_1f_2}(x_1,x_2,0)=D_{f_1}(x_1,0)\, D_{f_2}(x_2,0)\,
\frac{(1-x_1-x_2)^2}{(1-x_1)^{2+\alpha(f_1)}(1-x_2)^{2+\alpha(f_2)}}\;,
\ee
where $\alpha(f)=0$ for a sea quark and gluon and $\alpha(f)=0.5$ for valence quark, and with the fully factorized (FF) ansatz
(dot-dashed lines) 
\be\label{eq:44abc}
D_{f_1f_2}(x_1,x_2,0) = D_{f_1}(x_1,0)\, D_{f_2}(x_2,0)\,\theta(1-x_1-x_2)\;.
\ee
where the single PDFs are given by the MSTW08 parametrization \cite{Martin:2009iq}.  The main motivation for such  initial distributions   is the use of  single PDFs which are  well determined from global fits to data.  This allows  to constrain the DPDFs  which presently cannot be experimentally determined   due to the scarcity of data on double parton scattering. In both cases, condition (\ref{eq:32abc}) is 
only approximately fulfilled with 1.06 and 1.30 on the r.h.s  for the GS and FF prescriptions, respectively, which means that the momentum sum rule for the double parton  distributions is not exactly fulfilled. As a result, the small $x$ factorization, valid at the initial scale
(bottom left plot), is only approximate at the scale $Q^2=100~{\rm GeV}^2$ (right bottom plot). We see that the GS initial condition leads to a  slightly better factorization.  For a more detailed analysis of this ansatz, see  Ref.~\cite{Gaunt:2009re}.

The obvious drawback of the above initial conditions is the sum rules violation, which in principle makes the QCD evolution inconsistent. 
In Ref.~\cite{Golec-Biernat:2015aza} this problem was addressed in the pure gluonic case 
by taking advantage of a particular form  of the single PDFs in the MSTW08 parametrization. 
In principle, the entire construction can be repeated for the case with quarks, obtaining the double parton distributions satisfying both the momentum and valence quark number rules. However, such an  extension  requires additional constraints imposed on  the parameters of the single PDFs, which unfortunately  are not present in  realistic single PDFs obtained from fits to  data. Therefore, 
the GS proposal, which features factorization at small $x$, is at the moment the best practical way to
deal with the initial conditions for the DPDFs evolution which incorporates the knowledge about the single PDFs.

It should be emphasized at the end that the small $x$ factorization of the double parton distributions was discussed here only for the 
transverse momentum $\qbold=0$ when the sum rules are valid, see Section \ref{sec:2}.  For $\qbold\ne 0$, there
are  indications that the factorization does not hold due to correlations in the $\bbold-$space \cite{Blok:2011bu, Blok:2013bpa}. In particular, after the Fourier transformation to the $\qbold-$space,  the following ansatz  was proposed in \cite{Diehl:2014vaa} for $x_1,x_2<0.1$,
\be
D_{f_1f_2}(x_1,x_2,\qbold,0)=D_{f_1}(x_1,0)\, D_{f_2}(x_2,0)\exp\!\left\{-h_{f_1f_2}(x_1,x_2)\,\qbold^2\right\},
\ee
where $h_{f_1f_2}(x_1,x_2)$ is some function. Thus, for $\qbold=0$ the factorization (\ref{eq:44abc}) holds while for $\qbold\ne 0$ 
the factorization is always violated due to the exponential factor. This observation, however, does not invalidate our studies presented in this paper since the form of the double distrtibutions at $\qbold=0$ is essential for the generalization to the case with $\qbold\ne 0$.
For more details we refer to \cite{Diehl:2014vaa,Rinaldi:2016jvu, Diehl:2020xyg}.

\section{Summary}

We analyzed the role of the momentum sum rule  for the small
$x$ factorization of the double distributions into a product of single distributions in the pure gluonic case. In general, the momentum sum rule can be treated as consistency condition for the QCD evolution equations for single and double parton distributions. 
Therefore, it must be also fulfilled by initial conditions for these equations.

We found that  when the momentum sum rule is not fulfilled this can lead to strong violation of the small $x$ factorization at all scales.  On the other hand, when the momentum sum rule is imposed on the initial conditions,  then small $x$ factorization is  restored  to a good approximation  after the evolution to high scale. This was observed even when there was no factorization at the initial scale. We stress that for this to happen, the DPDF evolution equations must include  the splitting term.
 We also note that the small $x$ factorization observed at high scales is  approximate  with accuracy of a few percent. This indicates residual  correlations even at smallest values of $x$ and large values of $Q^2$.

We illustrated the impact  of  the momentum sum  rule on the issue of factorization on several numerical examples in the pure gluonic case and provide an analytic understanding  of the obtained results using the Mellin moment space. We also discussed the current status of the initial condition specifications for the QCD evolution,   which are motivated by a good knowledge of the single PDFs. We conclude that  the Gaunt-Stirling prescription for the initial double parton distributions leads to a phenomenologically acceptable prescription  for the small $x$ behavior of the double parton distributions
for the transverse momentum $\qbold=0$. However, it should be kept in mind that  the small $x$ factorization
can be violated for $\qbold\ne 0$ due to correlations in the $\bbold-$space.

\section{Acknowledgments}
This work was supported by the Polish Narodowe Centrum Nauki  Grant No.~2019/33/B/ST2/02588 and by  the U.S. Department of Energy Grant No.  DE-SC-0002145. We thank Michal Deak for the participation  in the early stages of this analysis.

\section{Appendix A: Momentum sum rule conservation}

We will prove that the momentum sum rule (\ref{eq:momrule1}) is conserved by the evolution equation (\ref{eq:twopdfeq})  by showing that the single PDFs obtained from Eq.~(\ref{eq:momrule1}),
\be
\label{eq:b1}
D_{f_2}(x_2,t)  = \frac{1}{1-x_2}\sum_{f_1}\int_{0}^{1-x_2}dx_1\,x_1D_{f_1f_2}(x_1,x_2,t) \; ,
\ee
obey the DGLAP equation (\ref{eq:dglapgl}). For this purpose we write the LO DGLAP equation in the most general form
\be
\label{eq:b2}
\partial_t D_f(x,t)= \sum_{\fprime} \int_0^1 du\, \Pcal_{f\fprime}(x,u)\, D_{\fprime}(u,t) \; ,
\ee
where $f$ indices denote quark/antiquark flavors and gluon. The splitting kernel has the following general form
\be
\label{eq:b3}
\Pcal_{f\fprime}(x,u)=P_{f\fprime}^R(x,u) - \delta(u-x)\,\delta_{f\fprime}\,P_{f}^V(x) \; ,
\ee
where $R$ and $V$ denote the real and virtual emission kernels, respectively. For the real emission kernel we  have
\be
\label{eq:b4}
P_{f\fprime}^R(x,u)= \frac{1}{u}P_{f\fprime}\!\!\left(\frac{x}{u}\right)\theta(u-x) \; ,
\ee
where $P_{f\fprime}$ are the LO real emission  Altarelli-Parisi splitting functions.
The virtual kernel is determined by assuming that the momentum sum rule for single PDFs is valid for any $t$,
\be
\sum_f \int_0^1dx\,  x D_f(x,t)= {\rm const} \; ,
\ee
which holds true during the DGLAP evolution if
\be
\label{eq:b5}
\sum_f \int_0^1dx\, x\,\Pcal_{f\fprime}(x,u)=0 \; ,
\ee
for any $f^\prime$ and $u\in[0,1]$. Substituting (\ref{eq:b3}) into the above we find
\be
u\,P_{\fprime}^V(u)=\sum_f\int_0^1 dx\, x\, P_{f\fprime}^R(x,u) \; .
\ee

In the introduced notation, the evolution equation (\ref{eq:twopdfeq}) for DPDFs is  given by
\begin{align}\nonumber
\label{eq:b7}
\partial_t D_{f_1f_2}(x_1,x_2,t)
&=  \sum_{\fprime}\int_{0}^{1-x_2}{du}\,\Pcal_{f_1\fprime}({x_1},{u}) \,
D_{\fprime f_2}(u,x_2,t)
\\\nonumber
&+\sum_{\fprime}\int_{0}^{1-x_1}{du}\,\Pcal_{f_2\fprime}({x_2},{u}) \,D_{f_1\fprime}(x_1,u,t)
\\
&+\sum_{\fprime}\, P_{\fprime\to f_1f_2} (x_1,x_2)\,D_{\fprime}(x_1+x_2,t)\;,
\end{align}
where the splitting kernel in the third term reads
\be
\label{eq:b8}
P_{\fprime\to f_1f_2}(x_1,x_2) = P^R_{f_1\fprime}(x_1,x_1+x_2) =P^R_{f_2\fprime}(x_2,x_1+x_2)\;.
\ee
The upper integration limits in Eq.~(\ref{eq:b7}) can be set to $1$ from the condition
\be\label{eq:b11a}
D_{f_1f_2}(x_1,x_2,t)=0~~~~~~{\rm for}~~~~~~x_1+x_2>1\;.
\ee
Under this condition the upper limit in Eq.~(\ref{eq:b1}) can also be set to $1$. Thus, after differentiating this equation,
\be
\partial_t D_{f_2}(x_2,t)  = \frac{1}{1-x_2}\sum_{f_1}\int_{0}^{1}dx_1\,x_1\,\partial_t D_{f_1f_2}(x_1,x_2,t)\;,
\ee
and using (\ref{eq:b7}) on the r.h.s., we obtain
\begin{align}
\label{eq:b11}
\nonumber
\partial_t D_{f_2}(x_2,t)  =
\sum_{f_1}\int_{0}^{1}dx_1\frac{x_1}{1-x_2}
&\bigg\{\sum_{\fprime}\int_{0}^{1}{du}\,\Pcal_{f_1\fprime}({x_1},{u})\,D_{\fprime f_2}(u,x_2,t)
\\\nonumber
&+\sum_{\fprime}\int_{0}^{1}{du}\,\Pcal_{f_2\fprime}({x_2},{u})\,D_{f_1 \fprime}(x_1,u,t)
\\
&+\sum_{\fprime}P_{\fprime\to f_1f_2} (x_1,x_2) \,D_{\fprime}(x_1+x_2,t)\bigg\}\;.
\end{align}
The first term on the rhs vanishes due to condition (\ref{eq:b5}). Thus, we find after changing the summation and integration order that
\begin{align}\nonumber
\label{eq:b12x}
\partial_t D_{f_2}(x_2,t)  &= \sum_\fprime \int_0^1\frac{du}{1-x_2}\,\Pcal_{f_2\fprime}({x_2},{u})
\bigg\{ \sum_{f_1} \int_0^1dx_1\,x_1 D_{f_1 \fprime}(x_1,u,t)\bigg\}
\\
&+\sum_\fprime \int_0^1 dx_1\,\frac{x_1}{1-x_2} \,P^R_{f_2\fprime}(x_2,x_1+x_2)\,D_{\fprime}(x_1+x_2,t)\;,
\end{align}
where we used  (\ref{eq:b8})  in the last equation  to write the second term in Eq.~(\ref{eq:b12x}). The sum over $f_1$ in this term disappears since for a given $(\fprime, f_2)$  there is only one $f_1=f_1(\fprime,f_2)$ in the sum. 
Applying  (\ref{eq:b1}) in the first term and changing the variable
$x_1\to u=x_1+x_2$  in the second one, we obtain
\be
\partial_t D_{f_2}(x_2,t)  = \sum_\fprime \int_0^1du\,\frac{1-u}{1-x_2}\,\Pcal_{f_2\fprime}({x_2},{u})\,D_\fprime(u,t)
+\sum_\fprime \int_{x_2}^{1+x_2} du\,\frac{u-x_2}{1-x_2} \,P^R_{f_2\fprime}(x_2,u)\,D_{\fprime}(u,t)\;.
\ee
Because of the property (\ref{eq:b4}) and $D_f(x,t)=0$  for $x>1$, 
the integration range in the second integral may be shifted to $[0,1]$.
Therefore, after taking into account the form (\ref{eq:b3}) of the kernel $\Pcal_{f_2\fprime}({x_2},{u})$ in the first integral, we find
\be
\partial_t D_{f_2}(x_2,t)  = \sum_\fprime \int_0^1du
\bigg\{
P^R_{f_2\fprime}({x_2},{u}) - \delta(u-x_2)\,\delta_{f_2\fprime}\,P^V_{f_2\fprime}({x_2})
\bigg\}D_{\fprime}(u,t)\;,
\ee
which is the DGLAP equation (\ref{eq:b2}). Notice the crucial role of the splitting term in the evolution equation (\ref{eq:b7}) to arrive at this conclusion.

\section{Appendix B: Asymptotic solution in Mellin moment space}

The solution to the evolution equation (\ref{eq:tevggl}) for the double gluon distribution in the Mellin space is given by 
\cite{Golec-Biernat:2014nsa}
\begin{align}
\tD_{gg}(n_1,n_2, t) &= \eto^{\gamma(n_1)t}\, \eto^{\gamma(n_2)t} \,\tD_{gg}(n_1,n_2,0) 
+\int_0^t dt' \eto^{\gamma(n_1)(t-t')}\,\eto^{\gamma(n_2)(t-t')}\, \tilde{\gamma}(n_1,n_2) \,\tD_g(n_1+n_2-1, t') \; .
\label{eq:dpdfsolm1}
\end{align}
The first term in the sum is the general solution to the  homogeneous equation (\ref{eq:tevggl}) without the splitting term while the second term  is a particular solution to the 
nonhomogeneous equation with the splitting term.

The homogeneous solution term  in Eq.~\eqref{eq:dpdfsolm1} contains a product of exponentials  
which generate two independent DGLAP evolutions in the double gluon distribution since
the solution to the DGLAP equation reads as
\be\label{eq:b2new}
\tD_g(n,t)=\eto^{\gamma(n) t}\,D_g(n,0) \; .
\ee
In  the nonhomogeneous solution term in  Eq.~\eqref{eq:dpdfsolm1}, the gluon splitting $g\to gg$ at "time"  $t^\prime$ is followed by two independent DGLAP evolutions with the exponential factors up to the final $t$. Since the splitting point $t^\prime$ might occur between
$0$ and $t$ we need to integrate over the whole range of such possibilities.

Let us concentrate  on the  nonhomogeneous solution in  Eq.~\eqref{eq:dpdfsolm1}, 
denoted from now on by $I$. Using (\ref{eq:b2new}), we obtain
\begin{align}\nonumber
\label{eq:b3new}
I &=\int_0^t dt' \eto^{\gamma(n_1)(t-t')+\gamma(n_2)(t-t')+\gamma(n_1+n_2-1)t'} \,\tilde{\gamma}(n_1,n_2)\, \tD_g(n_1+n_2-1,0) =
 \\
&= \eto^{\gamma(n_1)t +\gamma(n_2)t} \,  \tilde{\gamma}(n_1,n_2) \left(\frac{\eto^{(\gamma(n_1+n_2-1)-\gamma(n_1)-\gamma(n_2))t}-1}{\gamma(n_1+n_2-1)-\gamma(n_1)-\gamma(n_2)} \right)\tD_g(n_1+n_2-1,0)\; .
\end{align}
We will make the approximations  that reflect the small $x$ limit, in which the factorization is observed in the numerical analysis. To this aim, we observe that the  splitting function in this limit becomes:  ${{P}}_{gg}(z) \simeq 2 C_A /z$,  which leads to the following approximation to the anomalous dimension 
\be
\gamma(n)=\int_0^1 dz\,z^{n-1} \,{{P}}_{gg}(z) \simeq   \int_0^1 dz\,z^{n-1} \, \frac{2 C_A }{z} = \frac{2 C_A }{n-1} \; .
\ee
Thus, the small $x$ limit corresponds to the limit of $(n-1)\to 0$. For the double distribution we consider the small $x$ limit studied at the plots,
$x_1\to 0$ and $x_2={\rm fixed}$ and small, which corresponds in the Mellin moment space to the limit  $(n_1-1)\to 0$  and $(n_2-1)$ finite. 
In this limit, the function $\tilde{\gamma}$ given by Eq.~\eqref{eq:tdgamma}  is given by
\be
\tilde{\gamma}(n_1,n_2) \simeq 2 C_A \int_0^1dx\, x^{n_1-2} (1-x)^{n_2-1}=2 C_A \frac{\Gamma(n_1-1)\Gamma(n_2)}{\Gamma(n_1+n_2-1)}\simeq \frac{2 C_A }{n_1-1}\;,
\ee
while  for the expression in the denominator in Eq.~(\ref{eq:b3new}), we obtain
\be
\gamma(n_1+n_2-1)-\gamma(n_1)-\gamma(n_2) \; \simeq \;  -\frac{2 C_A }{n_1-1} \;.
\ee
So the nonhomogeneous solution (\ref{eq:b3new}) reads as
\begin{equation}
I\, \simeq \, e^{\gamma(n_1)t +\gamma(n_2)t}  \left[1-\exp\left(-\frac{2 C_A }{n_1-1}t\right)\right] \tD_g(n_1+n_2-1,0) \; .
\label{eq:approx1}
\end{equation}
 Here one needs to have ${\rm Re}(n_1)>1$ which is required for the inverse Mellin transform to make sense. Thus, 
for sufficiently large $t$, when the exponent in the square brackets can be neglected,  we find 
\be
I  \simeq \eto^{\gamma(n_1)t +\gamma(n_2)t} \tD_g(n_1+n_2-1,0) \; .
\ee
Therefore, the full solution  \eqref{eq:dpdfsolm1}   can be approximated by
\be
\label{eq:fact1}
\tD_{gg}(n_1,n_2,t) \simeq \eto^{\gamma(n_1)t +\gamma(n_2)t}\left( \tD_{gg}(n_1,n_2,0)+\tD_g(n_1+n_2-1,0)\right) \; .
\ee
This needs to be compared with the empirically found result in the considered small $x$ limit
\be
\label{eq:fact2}
\tD_{gg}(n_1,n_2,t) \simeq \eto^{\gamma(n_1)t +\gamma(n_2)t}  \; \tD_{g}(n_1,0)\tD_{g}(n_2,0) \;,
\ee
which gives the following relation for the Mellin moments of the initial condition to be satisfied for the small $x$ factorization
to be valid  at sufficiently large $t$, 
\be
\label{eq:b13new}
\tD_{gg}(n_1,n_2,0)+\tD_g(n_1+n_2-1,0)  = \tD_{g}(n_1,0)\,\tD_{g}(n_2,0) \; .
\ee


\bibliographystyle{h-physrev4}
\bibliography{mybib}

\end{document}